%% file: TOPQ-2017-15-PAPER.tex
\newcommand*{\ATLASLATEXPATH}{}
\documentclass[PAPER, atlasdraft=false, cernpreprint, texlive=2017, USenglish]{\ATLASLATEXPATH atlasdoc}
 
\usepackage{\ATLASLATEXPATH atlaspackage}
\usepackage{\ATLASLATEXPATH atlasbiblatex}
 
\usepackage{\ATLASLATEXPATH atlasphysics}
 
\graphicspath{{logos/}{figures/}}
 
\usepackage{TOPQ-2017-15-PAPER-defs}
 
\usepackage{multirow}

\AtlasTitle{Search for flavor-changing neutral currents in top quark
decays $t\to Hc$ and $t \to Hu$ in multilepton final states in proton--proton
collisions at $\sqrt{s}=$ 13~TeV with the ATLAS detector}

\AtlasAbstract{
Flavor-changing neutral currents are not present in the Standard Model at
tree level and are suppressed in loop processes by the unitarity of the
Cabibbo--Kobayashi--Maskawa
matrix; the corresponding rates for top quark decay processes are experimentally unobservable.  Extensions
of the Standard Model can generate new flavor-changing neutral current
processes, leading to signals which, if observed, would be unambiguous
evidence of new interactions. A dataset corresponding to an integrated luminosity of 36.1 fb$^{-1}$ of $pp$
collisions at a center-of-mass energy of $\sqrt{s}=$ 13~TeV recorded with the ATLAS detector at the
Large Hadron Collider is used
to search for top quarks decaying to up or charm quarks with the emission of
a Higgs boson, with subsequent Higgs boson decay to final states with at
least one electron or muon.  No signal is observed and limits on the
branching fractions  $\mathcal{B}(t \to Hc) <$
0.16\% and $\mathcal{B}(t \to H u) <$ 0.19\% at 95\% confidence level are
obtained (with expected limits of 0.15\% in both cases).
}

\AtlasRefCode{TOPQ-2017-15}
 
\PreprintIdNumber{CERN-EP-2018-067}

\AtlasJournal{Physical Review D}
\AtlasJournalRef{\PRD 98 (2018) 032002}
\AtlasDOI{10.1103/PhysRevD.98.032002}
 
\hypersetup{pdftitle={Search for flavor-changing neutral currents in top quark
decays in multilepton final states in proton--proton collisions at 13~\TeV\ with the ATLAS detector},pdfauthor={The ATLAS Collaboration}}
 
\begin{document}
 
\maketitle

\renewcommand{\arraystretch}{1.0}
 
\section{Introduction}
 
In the Standard Model (SM), the mass eigenstates in the quark sector couple
diagonally to the photon, \Zboson~boson, and Higgs boson, with the result that
quark flavors can only change at tree level by emission of \Wpm bosons (charged
currents).  Although processes that change quark flavors without external
emission of \Wpm\ bosons -- i.e.~flavor-changing neutral currents (FCNC) -- occur via loops in the SM, they are suppressed by the
Glashow--Iliopoulos--Maiani mechanism \cite{Glashow:1970gm}.  The decay
of a top quark to a Higgs boson and a lighter up-type quark $q$ ($t \to Hq$) is estimated to have
a branching fraction of about $3 \times 10^{-15}$ in the SM~\cite{AguilarSaavedra:2004wm}, which is unobservable with any current or
foreseeable dataset.  An observation of this process with current sensitivity
would be unambiguous evidence of physics beyond the Standard Model (BSM).
Searches for \tHq decays are part of a broader FCNC program that includes searches for $t\gamma q$, $tZq$, and
$tgq$ interactions \cite{CMS-TOP-14-003,TOPQ-2014-08,CMS-TOP-12-039,TOPQ-2014-13,CMS-TOP-14-007}.
 
Models of BSM physics can feature nontrivial flavor structures that produce
tree-level or large effective loop-induced $tHq$ couplings.  Tree-level
couplings are generic in two-Higgs-doublet models unless discrete symmetries
are introduced to forbid them \cite{Glashow:1976nt}, and can also be present in models with heavy
vector-like quarks \cite{delAguila:1985mk}.  The Cheng--Sher ansatz \cite{Cheng:1987rs} of off-diagonal
light Higgs boson interactions in models with multiple Higgs doublets gives
couplings that scale as $\lambda_{tHq} \sim \sqrt{2m_t m_q}/v$, where $m_t$ is
the top quark mass, $m_q$ is the light up-type quark mass, and $v$ is the
SM Higgs field vacuum expectation value. This would lead to a branching
fraction $\Br{t\to Hc}
\approx 0.15\%$, close to current experimental bounds. Smaller
loop-induced enhancements can be present in two-Higgs-doublet models even
without off-diagonal tree-level couplings \cite{Bejar:2000ub}, the minimal
supersymmetric Standard Model \cite{Guasch:1999jp}, $R$-parity-violating
supersymmetry \cite{Yang:1997dk}, models with
warped extra dimensions \cite{Agashe:2006wa}, and composite Higgs boson models~\cite{Agashe:2009di}. A summary of expectations for \tHq branching fractions
in various BSM models can be found in Ref.~\cite{Agashe:2013hma}.
 
The ATLAS and CMS collaborations have carried out searches
\cite{Aad:2015pja,Aad:2014dya,CMS-TOP-13-017,Aaboud:2017mfd,Sirunyan:2017uae}
for $tHq$ interactions with 7, 8,
and 13~\TeV\ $pp$ collision data from the Large Hadron Collider (LHC).  These
primarily searched for $t\bar t$ production where one top quark decays via $t \to
Wb$ and the other decays via \tHq; the analyses are distinguished by the
targeted Higgs boson decay.  The results at 13~\TeV\ benefit from the large
integrated luminosity delivered by the LHC in 2015--6 and from the \ttbar
cross section being larger at 13~\TeV\ than at 8~\TeV.  Using 13~\TeV\ data, ATLAS obtained
$\Br{\tHc} < 0.22\%$ with $H \to \gamma\gamma$ decays \cite{Aaboud:2017mfd} and CMS
obtained $\Br{\tHc} < 0.47\%$ with $H \to b\bar b$ decays \cite{Sirunyan:2017uae}, in both
cases at 95\% confidence level (CL) and assuming $\Br{\tHu} = 0$.  Similar
limits apply for $\Br{\tHu}$ assuming $\Br{\tHc} = 0$.
 
In this paper, a search for production of \ttbar pairs in which one top
quark decays via \tHq is reported,
targeting multilepton final states with either three light leptons ($\ell = e$ or $\mu$) or two
light leptons of the same electric charge.  The multilepton final states can be produced
via Higgs boson decays which involve light leptons in the final state through $H \to WW^*$, $H \to
\tau\tau$ in which both $\tau$-leptons decay leptonically, or $H \to ZZ^*$.  A \ttbar
pair with a \tHq decay, followed by $H \to WW^*$, would feature
the intermediate state $(Wb)(WW^*q)$, which can yield the final states $\ell\ell
b 3q 2\nu$ or $\ell\ell\ell b q 3\nu$ ($q$ representing any quark lighter than the $b$-quark).
Events with hadronically decaying $\tau$-lepton candidates are vetoed to avoid overlap
with dedicated searches for those Higgs boson decay modes.
 
The sought-after \ttbar, \tHq signature with Higgs boson decays into final
states with leptons is in many respects
similar to the corresponding channel for \ttH production, although \tHq has
one fewer $b$-jet and one fewer light-quark jet for the same lepton multiplicity in the
absence of additional radiation.  The sample
of events and background simulations used for the 13~\TeV\ ATLAS search for \ttH production in the multilepton final state~\cite{Aaboud:2017jvq} is therefore
leveraged in this search, similar to the strategy used in the 8~\TeV\
multilepton \tHq
search~\cite{Aad:2015pja}.  Unlike the 8~\TeV\ search, however, in this
analysis multivariate discriminants (boosted decision trees, BDTs) are used,
optimized to separate the FCNC signal from SM processes.
 
The analysis proceeds as follows.  The same-charge dilepton and
trilepton categories (with no hadronic $\tau$-lepton candidates) from the 13~\TeV\ \ttH search \cite{Aaboud:2017jvq}
are used.  The same event preselection, calibration, and SM simulation samples are used as in
the \ttH search, but the \ttH process is now treated as a background and
fixed to the expected SM rate.  The FCNC processes \tHu and \tHc are
simulated and BDTs are trained to separate these from SM processes in the two
categories.  Care is taken to account for FCNC signal contamination in
control regions which are used to constrain backgrounds arising from lepton production
in hadron decays and photon conversions.  The results are obtained by fitting the data
distributions of the BDT discriminants.

\section{ATLAS detector and object reconstruction}
\label{sec:detector}
 
\newcommand{\AtlasCoordFootnote}{
ATLAS uses a right-handed coordinate system with its origin at the nominal interaction point (IP)
in the center of the detector and the $z$-axis along the beam pipe.
The $x$-axis points from the IP to the center of the LHC ring,
and the $y$-axis points upwards.
Cylindrical coordinates $(r,\phi)$ are used in the transverse plane,
$\phi$ being the azimuthal angle around the $z$-axis.
The pseudorapidity is defined in terms of the polar angle $\theta$ as $\eta = -\ln \tan(\theta/2)$.
Angular distance is measured in units of $\Delta R \equiv \sqrt{(\Delta\eta)^{2} + (\Delta\phi)^{2}}$.}

The ATLAS experiment~\cite{PERF-2007-01} at the LHC is a multipurpose particle detector
with a forward-backward symmetric cylindrical geometry and near $4\pi$ coverage in
solid angle.\footnote{\AtlasCoordFootnote}
It consists of an inner tracking detector surrounded by a thin superconducting solenoid
providing a \SI{2}{\tesla} axial magnetic field, electromagnetic and hadron calorimeters, and a muon spectrometer.
The inner tracking detector covers the pseudorapidity range $|\eta| < 2.5$.
It consists of silicon pixel, silicon microstrip, and transition radiation
tracking detectors. A new innermost silicon pixel layer was installed prior to
data-taking at $\sqrt{s} = 13$~\TeV~\cite{CERN-LHCC-2010-013}.
Lead/liquid-argon (LAr) sampling calorimeters provide electromagnetic (EM) energy measurements
with high granularity for $|\eta| < 3.2$.
A hadron steel/scintillator-tile calorimeter covers the central pseudorapidity range ($|\eta| < 1.7$).
The endcap and forward regions are instrumented with LAr calorimeters
for both the EM and hadronic energy measurements up to $|\eta| = 4.9$.
The muon spectrometer surrounds the calorimeters and is based on
three large air-core toroidal superconducting magnets with eight coils each.
The muon spectrometer includes a system of precision tracking chambers and fast detectors for triggering.
A combined hardware and software trigger system is used to select events.
The first-level trigger is implemented in hardware and uses a subset of the detector information
to reduce the accepted rate to a design maximum of \SI{100}{\kilo\hertz}.
This is followed by a software-based trigger that
further reduces the accepted event rate to a sustained
rate of about \SI{1}{\kilo\hertz}.
 
This analysis uses an integrated luminosity of 36.1~\ifb\ of $pp$ collision data with $\sqrt{s} = 13$~\TeV\
collected during 2015 and 2016, corresponding to approximately 30 million
$t\bar t$ pair production events \cite{Czakon:2011xx}.  The mean number of
$pp$ interactions per bunch crossing in the dataset is 24. A full description of the reconstruction and selection of physics objects can be found in
Ref.~\cite{Aaboud:2017jvq}, and only a brief description follows.
 
Events in this analysis are selected using triggers which require the
presence of one or two light leptons \cite{Trigger2015}.  The single-lepton triggers have transverse momentum (\pt)
thresholds that vary from 20 to 26~\GeV\ depending on lepton flavor and instantaneous luminosity.  In the dilepton triggers the
thresholds for the higher-\pt lepton vary from 12 to 22~\GeV\ depending on lepton flavor and instantaneous luminosity.
 
Muon candidates are formed using inner detector tracks and muon spectrometer
tracks, track segments, or for $|\eta| < 0.1$, calorimeter signals consistent
with the passage of a minimum-ionizing particle.  They are required to satisfy $\pt > 10$~\GeV\ and $|\eta| < 2.5$ and to pass \textit{loose} identification requirements \cite{Aad:2016jkr}, as
well as transverse
and longitudinal track impact
parameter requirements with respect to the primary vertex. The primary vertex
in an event is defined as the
reconstructed $pp$ collision vertex with the highest $\sum \pt^2$ of
associated tracks with $\pt > 400$~\MeV.
 
Electron candidates are formed from energy clusters in
the electromagnetic calorimeter associated with inner detector tracks. They are
required to satisfy $\pt > 10$~\GeV\ and $|\eta_\mathrm{cluster}| < 2.47$,
excluding the transition region $1.37 < |\eta_\mathrm{cluster}| < 1.52$
between the barrel and endcap electromagnetic calorimeters.  A likelihood
discriminant is used to separate isolated prompt electrons from hadronic jets, photon
conversions, and nonisolated electrons from hadron decays; two working points
of the discriminant
(\textit{loose} and \textit{tight} \cite{Aaboud:2016vfy}) are used in this analysis.  Electron candidates must also
pass transverse and longitudinal track impact parameter requirements.
 
Jets are reconstructed using clusters built from energy deposits in the
calorimeters \cite{PERF-2014-07,ATL-PHYS-PUB-2015-036,PERF-2016-04} using the \antikt algorithm
\cite{Cacciari:2008gp,Cacciari:2011ma} with radius parameter $R = 0.4$.
Considered jets are required to satisfy $\pt > 25$~\GeV\ and $|\eta| < 2.5$.
Jets with $\pt < 60$~\GeV\ and $|\eta| < 2.4$ are required to be matched to the
primary vertex using their associated inner detector tracks \cite{PERF-2014-03}.  Jets containing
$b$-hadrons ($b$-jets) are tagged using a multivariate discriminant
combining information about secondary vertices, reconstructed decay chains, and impact parameters of tracks
associated with the jet \cite{PERF-2012-04,ATL-PHYS-PUB-2016-012}.  The working point for $b$-jet identification used in
this analysis corresponds to an average efficiency of 70\% for jets containing
$b$-hadrons with $\pt > 20$~\GeV\ and $|\eta| < 2.5$ in simulated \ttbar events.  Jets
which contain $c$-hadrons but not $b$-hadrons have approximately a one in
twelve probability of being misidentified as $b$-jets; the same probability for light-quark or gluon jets is one in 380, leading to different
response for the \tHu and \tHc processes.
 
Hadronically decaying $\tau$-lepton (\tauhad) candidates are reconstructed from hadronic jets associated with either one or three
inner detector tracks with a total charge of $\pm 1$ \cite{ATL-PHYS-PUB-2015-045,ATLAS-CONF-2017-029}. A BDT discriminant is
used to distinguish \tauhad candidates from quark- or gluon-initiated jets.  A
working point with 55\% (40\%) efficiency for one-prong (three-prong) \tauhad
decays is used for the veto. Candidates with $\pt > 25$~\GeV\ and $|\eta| < 2.5$ are considered.
 
The missing transverse momentum, with magnitude \met, is defined as the
negative vector sum of the transverse momenta of all calibrated and identified
leptons and jets and the remaining unclustered energy of the event, which is estimated from low-\pt tracks associated with the primary vertex
but not with any considered lepton or jet \cite{Aaboud:2018tkc}.
 
To eliminate ambiguity between reconstructed objects and reduce the fraction of leptons
arising from hadron decay, the following additional requirements are imposed:
if two electrons are separated by $\DeltaR < 0.1$, only the higher-\pt
electron is considered; any electron within $\DeltaR = 0.1$ of a selected muon
is rejected; jets within $\DeltaR = 0.3$ of a selected electron are removed;
any \tauhad candidate within $\DeltaR = 0.2$ of a selected electron or muon is
ignored;
and muons must be separated by $\DeltaR > \mathrm{min}(0.4, 0.04 + (10\
\GeV)/p_{\mathrm{T},\mu})$ from a jet surviving the above selection.
 
To further reject backgrounds, BDTs are trained to discriminate against
electrons arising from asymmetric trident processes $e^\pm \to e^\pm e^+e^-$
in detector material
which may induce an apparent change of the electron charge, and against leptons
which arise from decays of $b$-hadrons \cite{Aaboud:2017jvq}.  The
former discriminant (``charge misassignment BDT'') uses track and calorimeter cluster information; the
latter (``non-prompt lepton BDT'') uses information about additional particle
activity and properties of low-\pt jets formed from tracks
near the lepton, including the output of $b$-tagging algorithms run on those
jets.  Working points with efficiencies of 95\% (60--98\%) are defined for the
charge misassignment (non-prompt lepton) discriminants.
 
During the initial loose (L) selection of leptons, no isolation, charge
misassignment discriminant, or non-prompt lepton discriminant requirements are
imposed. Electrons are required to pass the loose electron identification
discriminant selection. Two other lepton selections are used in the analysis:
\begin{itemize}
\item Modified loose (L*), in which the lepton must pass calorimeter- and
track-based isolation criteria and the non-prompt lepton
discriminant selection.
\item Tight (T), in which electrons must pass the tight electron
identification discriminant selection and the charge misassignment
discriminant selection. For muons the T and L* selections are
identical; this corresponds to the T* selection of Ref.~\cite{Aaboud:2017jvq}.
\end{itemize}

\section{Simulation, event selection, and analysis}
 
 
The simulated $pp \to \ttbar$, \tHq signals were generated with next-to-leading-order (NLO) QCD matrix
elements computed by \textsc{MadGraph5\_aMC@NLO} \cite{Alwall:2014hca}, with top quark decays performed by
MadSpin \cite{Artoisenet:2012st}; \textsc{Pythia 8} \cite{Sjostrand:2007gs} was used for Higgs boson decay, parton
showering, hadronization, and underlying-event generation. Either the top quark or antiquark may undergo the FCNC decay in this sample.  The total
top quark pair production cross section used to normalize the FCNC signal was set to $832 ^{+40}_{-46}$ pb, as
calculated with the Top++2.0 program at next-to-next-to-leading order in
perturbative QCD, including soft-gluon resummation to
next-to-next-to-leading-log order~\cite{Czakon:2011xx}.  The systematic
uncertainties include variation of factorization and renormalization scales
as well as uncertainties in parton distribution functions (PDFs) and the QCD coupling
$\alpha_\text{s}$~\cite{Botje:2011sn,Martin:2009bu,Gao:2013xoa,Ball:2012cx}.
 
The simulations of
SM background processes are the same as those used in
Ref.~\cite{Aaboud:2017jvq}. In particular, the major processes \ttZ, \ttW, and
\ttH were generated at NLO in QCD
with \textsc{MadGraph5\_aMC@NLO} interfaced to \textsc{Pythia
8} for
parton showering, hadronization, particle decay, and underlying-event generation.
The top quark and
Higgs boson masses were set to 172.5 and 125.0~\GeV, respectively.  Higgs boson
decay branching fractions were taken from Ref.~\cite{deFlorian:2016spz}. In all the
preceding samples, the matrix element calculations used the NNPDF 3.0 NLO
PDF set \cite{Ball:2014uwa}, while the parton shower calculations used the
A14 tune of \textsc{Pythia 8} parameters \cite{ATLASUETune4} and the NNPDF
2.3 LO PDF set \cite{Ball:2012cx}.  Diboson production was generated at NLO in
QCD with \textsc{Sherpa} \cite{sherpa,Hoeche:2012yf}
using the CT10 PDF set \cite{ct10}.  The response
of the ATLAS detector to generated events was simulated with
a \textsc{Geant4}-based detector model \cite{geant4,SOFT-2010-01} with parameterization of the calorimeter
response for some minor backgrounds \cite{ATL-PHYS-PUB-2010-013}.
 
The two categories of events used in this analysis, same-charge dilepton
(\twol) and trilepton (\threel) candidates, are selected by the same
requirements as in
Ref.~\cite{Aaboud:2017jvq}.  In both cases the leptons identified by the
trigger must correspond to leptons selected for this analysis, with
sufficiently high \pt for the trigger to be fully efficient.
\begin{itemize}
\item Events in the \twol category must have at least four reconstructed jets,
of which one or two must be $b$-tagged jets.  Exactly two light-lepton candidates
meeting L criteria (as described in Section~\ref{sec:detector})
must be found, along with no \tauhad candidates.  The leptons are then
also required to pass the T selection.  Both leptons must have $\pt > 20$~\GeV\ and have the same charge.
\item Events in the \threel category must have at least two reconstructed
jets, of which at least one must be a $b$-tagged jet.  Exactly three light-lepton candidates
meeting L criteria must be found, along with no \tauhad candidates.  The total
charge of the leptons must be $\pm 1$.  Of the three leptons, one
(designated $\ell_0$) is of opposite charge to the other two, and the
other two leptons are designated $\ell_1$ and $\ell_2$ in order of
increasing angular separation \DeltaR from $\ell_0$.  The two leptons $\ell_1$
and $\ell_2$ are required to meet the T criteria and have $\pt > 15$~\GeV.
The remaining lepton, $\ell_0$, is less likely to be non-prompt than either of the same-charge leptons $\ell_1$ or $\ell_2$.
It is correspondingly only required to meet the L* criteria and to have $\pt > 10$~\GeV.  To remove
contamination from hadron decay chains including $\ell^+\ell'^-$, both
invariant masses $m(\ell_0
\ell_1)$ and $m(\ell_0 \ell_2)$ must exceed 12~\GeV.  To remove
contamination from \ttZ, a $Z$ boson veto is imposed: $|m(\ell^+ \ell^-) -
91.2\ \GeV| > 10\ \GeV$ for every opposite-charge lepton pair of the same
flavor ($e^+e^-$ or $\mu^+ \mu^-$).  Finally, contamination from $\Zboson
\to \ell\ell\gamma^{(*)} \to \ell\ell \ell'(\ell')$, where one lepton has
low momentum and is not reconstructed, is removed by requiring
$|m(3\ell) - 91.2\ \GeV| > 10\ \GeV$.
\end{itemize}
After these selections, the \tHq signal is dominated by $H \to WW^*$ (85\% of
the \twol and 71\% of the \threel category) with subleading contributions
from $H \to \tau \tau$ (12\% and 16\% respectively) and $H \to ZZ^*$ (2\% and
9\% respectively). The fraction of \ttbar events with a consequent \tHq decay which are
expected to be reconstructed and selected is $5.1 \times 10^{-4}$ ($2.6 \times 10^{-4}$) for the \twol
(\threel) category.\footnote{These values include the effects of selection
acceptance, detector efficiency, and decay branching fractions.}
 
Following the initial selections, the largest sources of background are those
arising from non-prompt leptons (from hadron decays, photon conversions, and
charge misassignment), mainly from \ttbar decays, and prompt lepton
backgrounds from \ttV production ($V=W$ or $Z$) with leptonic decays of the
vector boson.  Further BDT
discriminants are trained to separate the FCNC signal from these two
background sources.
 
Inputs to the BDTs include lepton flavor and kinematic observables, jet properties
including whether they are $b$-tagged, angular separations between objects, the \met,
and the quantity $m_\mathrm{eff} \equiv \met + H_\mathrm{T}$, where
$H_\mathrm{T}$ is the scalar sum of the \pt of leptons and jets in the
event. Signal events can
be distinguished from \ttV and non-prompt lepton background by having only one true $b$-jet and being relatively soft events
with low $m_\mathrm{eff}$ and $H_\mathrm{T}$. The spin correlation in the dominant $H \to WW^*$
decay, and the presence of an off-mass-shell $W^*$ boson, also
yields a distinct signature where, in both categories, one lepton often has
low \pt, and in the \threel category, both $m(\ell_0\ell_1)$ and $\DeltaR(\ell_0,\ell_1)$ are small.
 
The variables used in the training of the BDTs in the two categories are shown
in Table~\ref{tbl:bdtvars}, and example distributions are shown in
Figure~\ref{fig:illustration}.  Good agreement is observed between the data
and expected background distributions in each variable.  In the \twol
category, the variables that most strongly separate \tHq from the non-prompt lepton background
are the number of $b$-jets, the \pt of the
lower-\pt lepton, and the angular separation \DeltaR of the lower-\pt lepton
from the closest jet.  The separation from \ttV processes comes mainly from the number
of $b$-jets, $m_\mathrm{eff}$, \met, and the \pt of the higher-\pt lepton.  In the
\threel category the strongest separation from non-prompt lepton backgrounds
comes from $\DeltaR(\ell_0, \ell_1)$, the invariant masses $m(\ell_0, \ell_1)$
and $m(\ell_0, \ell_2)$,
the \pt of the highest-\pt lepton, and the invariant mass of the opposite-charge, same-flavor lepton pair with mass closest to that of the $Z$ boson.  The
strongest separation from \ttV comes from the invariant masses $m(\ell_0, \ell_1)$
and $m(\ell_0, \ell_2)$, $m_\mathrm{eff}$, the number of $b$-jets, and the invariant
mass of the three leptons.  In both the \twol and \threel categories and
against both backgrounds, better
separation is achieved for \tHu than \tHc signals, as the latter is more
likely to have a second $b$-tagged jet arising from the hadronization products
of the charm quark. In the \twol category, \tHu and \tHc are sufficiently
similar that only one discriminant is trained for
the two decay modes.  In the \threel category the two signals
are treated separately.
 
\begin{table}
\begin{center}
\caption{\label{tbl:bdtvars}Variables used to construct the BDT
discriminants for the \twol and \threel categories.  The symbol
``$\times$'' indicates that this variable is used in the respective BDT. The ``best \Zboson
candidate'' is the opposite-charge lepton pair with same flavor with
invariant mass closest to 91.2~\GeV; if no such pair exists, zero is
assigned for the invariant mass.}
\begin{tabular}{lcc}
\hline\hline
Variable & \twol & \threel\\
\hline
\pt of higher-\pt lepton & $\times$ & \\
\pt of lower-\pt lepton & $\times$ & \\
\pt of lepton $\ell_0$ & & $\times$\\
\pt of lepton $\ell_1$ & & $\times$\\
\pt of lepton $\ell_2$ & & $\times$\\
Dilepton invariant masses (all combinations) & $\times$ & $\times$\\
Trilepton invariant mass & & $\times$ \\
Best $Z$ candidate invariant mass & & $\times$ \\
Maximum lepton $|\eta|$ & $\times$ & \\
Lepton flavor & $\times$ & \\
Number of jets & $\times$ & $\times$\\
Number of $b$-tagged jets & $\times$ & $\times$\\
\pt of highest-\pt jet & & $\times$\\
\pt of second highest-\pt jet & & $\times$\\
\pt of highest-\pt $b$-tagged jet & & $\times$\\
$\DeltaR(\ell_0, \ell_1)$ & & $\times$\\
$\DeltaR(\ell_0, \ell_2)$ & & $\times$\\
$\DeltaR(\textrm{higher-}\pt\textrm{ lepton}, \textrm{closest jet})$ & $\times$ & \\
$\DeltaR(\textrm{lower-}\pt\textrm{ lepton}, \textrm{closest jet})$ & $\times$ & \\
$\DeltaR(\ell_1, \textrm{closest
jet})$ & & $\times$\\
Smallest $\DeltaR(\ell_0, b\textrm{-tagged jet})$ & & $\times$  \\
\met & $\times$ & \\
$m_\mathrm{eff}$ & $\times$ & $\times$\\
\hline\hline
\end{tabular}
\end{center}
\end{table}
 
\begin{figure}
\begin{center}
\includegraphics[width=.48\linewidth]{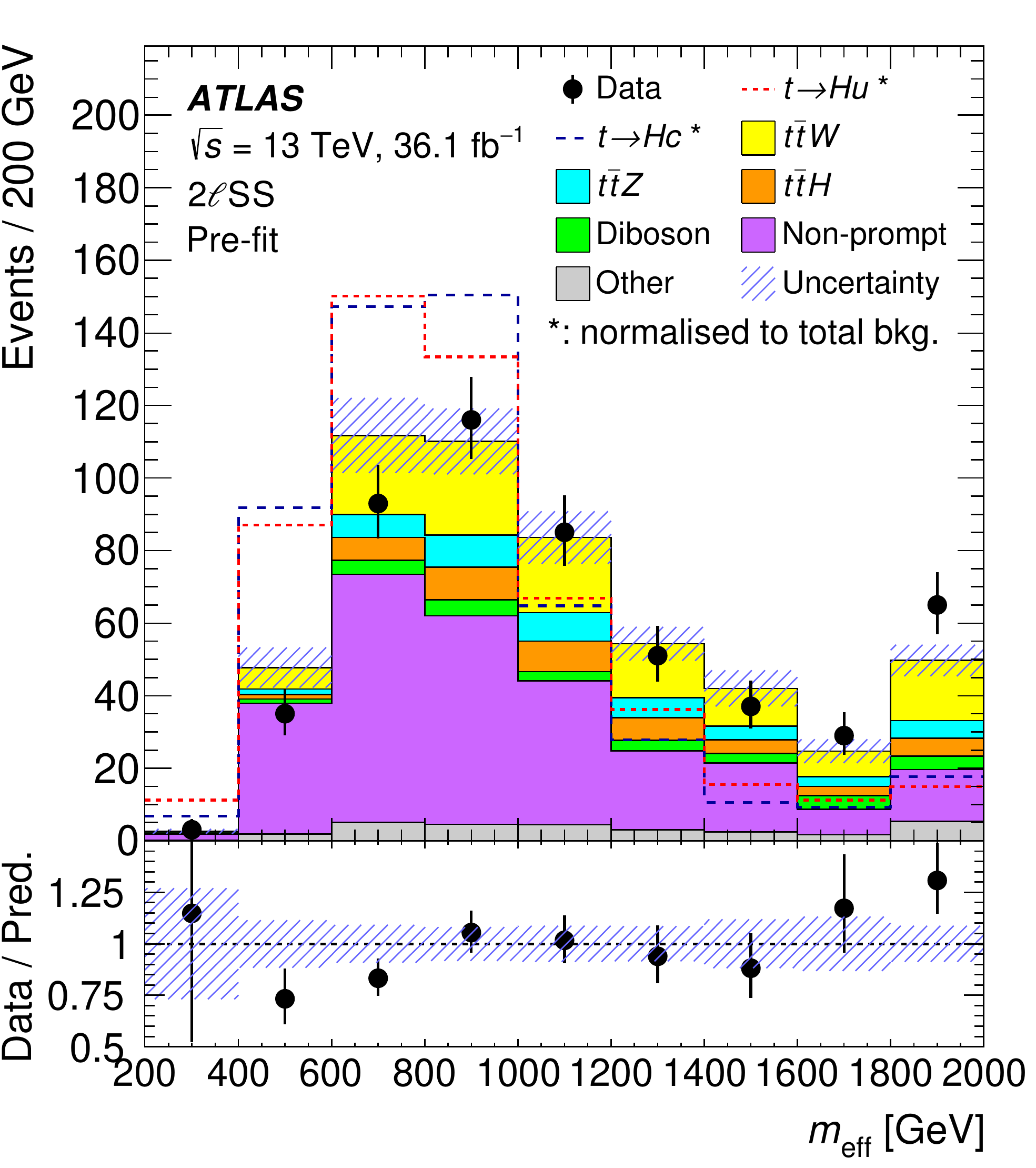}
\hfill\includegraphics[width=.48\linewidth]{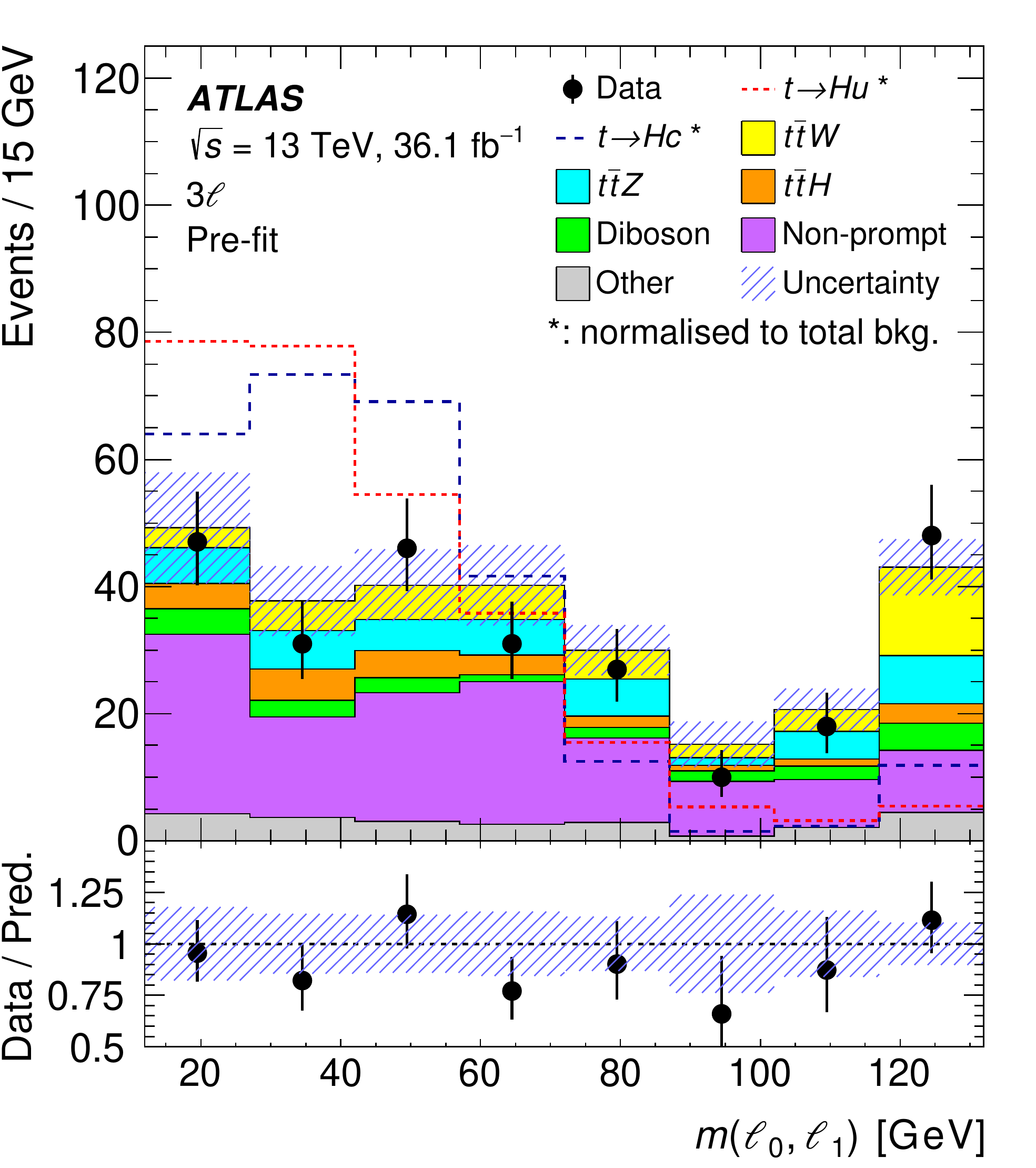}
\caption{\label{fig:illustration}Examples of distributions of variables separating FCNC signals
from SM background: (left) the quantity \meff\ (defined in the text) in the \twol category; (right) the
invariant mass $m(\ell_0, \ell_1)$ of the opposite-charge lepton pair with the smaller angular
separation \DeltaR in the \threel category. The bottom panels show
the ratio of the observed data yields in each bin to the SM prediction. The hashed bands indicate the
total uncertainty in the SM prediction.  The FCNC signal distributions are
shown as open histograms, normalized to the same total yield as the SM
backgrounds. The rightmost bins include overflow. }
\end{center}
\end{figure}
 
In each case, two separate discriminants are trained: one to separate \tHq
from non-prompt leptons and one to separate \tHq from \ttV processes.  The
former is designated BDT$(\ttbar, X)$ and the latter BDT$(\ttV, X)$ for a
given category/signal choice $X$ (i.e., \twol, \threel \tHu, or \threel
\tHc).  The expected distributions for backgrounds with non-prompt leptons are
obtained using the procedures described in Section~\ref{sec:fakes}.
The two discriminants are combined into a single discriminant,
designated BDT$(X)$, via a linear combination that yields the best expected
limit on the FCNC branching fraction.
 
The BDT discriminant outputs in
the \twol and \threel categories are binned into six or four bins,
respectively, with bin widths optimized to provide the best expected limits.
Every bin contains a roughly equal number of signal events.  The bin
boundaries used for the fits to \tHu and \tHc signals in the \twol category are
optimized separately, although the discriminant used is the same for the two decays.

\section{Background estimation}
\label{sec:fakes}
The estimation of rates and kinematic distributions of SM processes that
form backgrounds with prompt leptons in the signal categories
is performed using simulation.  The processes considered include:
\begin{itemize}
\item $\ttbar W$, $\ttbar (Z/\gamma^* \to \ell\ell)$, $\ttbar H$, and $\ttbar
WW$;
\item $t\bar t t$ and $\ttbar\ttbar$;
\item single top quark production in the s- and t-channels, $tW$, $tZ$, $tWZ$,
$tHb$, and $tHW$;
\item production of two or three $W$ or $Z/\gamma^*$ bosons.
\end{itemize}
Details of the simulations used are given in Ref.~\cite{Aaboud:2017jvq}.
 
The estimation of the non-prompt lepton backgrounds, including the
contribution from
charge-misassigned electrons, uses data-driven methods
following Ref.~\cite{Aaboud:2017jvq}.  One major
modification to the treatment of leptons from hadron decays and conversions
is made, arising from the lower expected jet multiplicity in \tHq events
compared to \ttH production.  A summary of
the procedure is given below.  The self-consistency of the methods is checked
by predicting the non-prompt lepton yields in simulated SM \ttbar and
$\ttbar \gamma$ events; good agreement is observed.
 
Rates for electron charge
misassignment as functions of electron \pt and $|\eta|$ are determined from a
sample of $Z \to ee$ events where both electrons are reconstructed with the
same charge.
The contribution of these events to the \twol signal region is
determined by applying these misassignment rates to events selected with the
\twol selection criteria except requiring opposite-charge leptons instead of
same-charge leptons.
 
The estimation of non-prompt lepton background contributions other than
electron charge misassignment is performed using the so-called \textit{matrix method} \cite{Abbott:1999tt,ATLAS-CONF-2014-058}. This
technique uses control regions with the same kinematic properties as the signal region,
but with changed lepton identification criteria that enhance the non-prompt lepton
contribution, to statistically estimate the fraction of signal region events
that involve non-prompt leptons. Lepton candidates that meet L, but not
T, identification criteria ($\mathrm{\bar T}$) are much more likely to be of non-prompt
origin than those that meet T criteria; if the probabilities (``efficiencies'') of
prompt and non-prompt leptons to be identified as $\mathrm{\bar T}$ or T are
known, the fraction of tight-lepton events of non-prompt origin can be
determined by solving a system of equations.
The matrix
method can be applied to estimate contributions arising from multiple non-prompt leptons in
the same event.  In this analysis, the two leptons in the \twol category
events, and the two same-charge leptons in the \threel category events, are
analyzed for a non-prompt contribution.  The opposite-sign
lepton $\ell_0$ in \threel events is found in simulation to be prompt 97\% of the
time, so the matrix method is not applied to that lepton.
 
The non-prompt efficiencies are measured in control regions with the same
selection criteria as the \twol signal category except that the number of jets must
be two or three, one lepton need only satisfy L criteria, and the lower-\pt
lepton is only required to have $\pt > 15$~\GeV.  These are then separated
into TT or T$\mathrm{\bar T}$ events. The expected number of events from SM
processes with only prompt leptons in these regions is
determined from simulation and subtracted from the observed number of events,
giving a yield of non-prompt lepton events which is then used to determine the
non-prompt efficiencies.  In the case of a nonzero \tHq branching fraction, a significant
fraction of the signal will be reconstructed in these control regions.  This
will act as an additional source of prompt leptons which is not accounted for
in the SM prediction and will bias the nominal
efficiencies which are determined assuming zero signal contribution.  For
$\Br{\tHq} = 0.2\%$, the FCNC process would contribute approximately 30\% of
the prompt lepton contribution in the low-jet-multiplicity control regions
with T leptons.  This
effect is accounted for in two ways.  First, the non-prompt efficiencies are
derived under the two hypotheses $\Br{\tHq} = 0$ and $\Br{\tHq} = 0.2\%$ and
both values are used to predict the yield of non-prompt leptons in the signal
categories; the two hypotheses result in non-prompt yields differing by $\approx$
40\% for the \twol and $\approx$ 30\% for the \threel category.  This overall normalization
correction from possible signal contamination is then scaled proportionally to the FCNC branching fraction
in the fit. Second, the
change in the shape of the FCNC discriminant response for the non-prompt
background in the signal regions under the two hypotheses is derived. The
difference is assigned as a systematic uncertainty on the
non-prompt background discriminant shape.
 
The separation of FCNC signal from non-prompt lepton
background by the BDT($\ttbar, X$) discriminants is sufficiently strong that the impact of these systematic uncertainties in the non-prompt
background estimate on the signal extraction is small.  Tests with
MC simulation indicate that the procedure correctly recovers the branching fraction of injected FCNC signals.

 
\section{Systematic uncertainties}
 
The same model of systematic uncertainties in background processes (including \ttH) is used as
in Ref.~\cite{Aaboud:2017jvq}, with the additional normalization and BDT shape
uncertainties in non-prompt
lepton backgrounds described in Section~\ref{sec:fakes}.
As the measured \ttH
cross section is compatible with the SM predictions, the SM rate is assumed
with appropriate theoretical uncertainties.  Acceptance uncertainties from the choice of
parton distribution functions and QCD scale for the major backgrounds
simulated with \textsc{MadGraph5\_aMC@NLO} are calculated using
SysCalc \cite{Kalogeropoulos:2018cke}.
 
The \tHq signal processes are subject to their own theoretical uncertainties,
primarily in the modeling of the parent \ttbar system.  Systematic
uncertainties are assigned for the \ttbar cross section, the variation of BDT
response with the choice of renormalization and factorization scale, the modeling of
parton showers, the event generator, and the amount of initial/final-state radiation.
 
The systematic uncertainty model includes components from:
\begin{itemize}
\item light lepton, \tauhad, and jet selection and energy/momentum
scale and \met modeling;
\item $b$-jet tagging efficiency and the probability for $c$-jets
and light-quark or gluon jets to be misidentified as $b$-jets;
\item the cross section and MC modeling of
simulated backgrounds and signals;
\item the statistical uncertainties in the control regions
for non-prompt lepton backgrounds, the matrix method efficiencies, and the
applicability to the \twol and \threel category events of the matrix method efficiencies derived at low jet
multiplicity in same-charge dilepton events;
\item electron charge misassignment;
\item $pp$ integrated luminosity (determined using a methodology
similar to that described in Ref.~\cite{DAPR-2013-01});
\item modeling of multiple $pp$ interactions per bunch crossing.
\end{itemize}
 
The background-related systematic uncertainties with the largest impact on the final result are
found to be those associated with the statistical uncertainty in the non-prompt lepton background estimation, the non-prompt lepton
efficiencies used in the matrix method, and the cross section for diboson
production in association with $b$-quarks.  Excluding the correction for
signal contributions to non-prompt lepton control regions, systematic
uncertainties on the background lead to an uncertainty
in the determined signal decay branching fractions of 0.04\%. Systematic uncertainties in the signal processes are primarily associated
with the matching of matrix element calculations with parton shower algorithms
and different choices of parton shower algorithms. The relative systematic
uncertainty in the signal yield prediction for a given signal branching fraction is 8\%.  The uncertainty in the background estimate due to the signal contributions in non-prompt lepton control regions depends on the signal decay branching fraction; for a true branching fraction of 0.2\%, the corresponding systematic uncertainty on the determined branching fraction is 0.02\%.
 
\section{Results}
\begin{table}
\begin{center}
\caption{\label{tbl:yields}Expected SM background (including
non-prompt leptons), FCNC contributions, and observed data yields in the
signal categories. The FCNC contribution shown in the ``pre-fit'' rows is the
expected yield for a 0.2\% branching fraction, and the result in the
``post-fit'' rows is the best-fit yield from the
combined fit of the \twol and \threel categories.  The uncertainties shown for the FCNC
yields reflect systematic uncertainties given a specific branching fraction
(pre-fit) or the uncertainties in the yield from the full fit (post-fit). SM backgrounds are determined as
described in Section~\ref{sec:fakes}.  The non-prompt lepton background
estimate includes electron charge misassignment. The non-prompt lepton component of
the pre-fit SM background is determined assuming zero FCNC branching fraction;
the post-fit non-prompt lepton background yield includes the effect of nonzero
FCNC branching fraction.}
{\small
\begin{tabular}{clcccccccc}
\hline\hline
\multicolumn{2}{l}{Category} & Non-prompt & \ttV & \ttH & Diboson & Other &
Total SM & FCNC & Data \\
& & leptons & & & & prompt SM & & &\\
\hline
\multicolumn{10}{c}{\tHu}\\
\hline
\multirow{2}{*}{\twol}
& Pre-fit & $266 \pm 40$ & $165 \pm 19$ & $43 \pm 4$ & $25 \pm 15$ & $28 \pm
6$ & $526 \pm 39$ & $61 \pm 13$
& \multirow{2}{*}{514}\\
& Post-fit & $240 \pm 37$ & $167 \pm 18$ & $43 \pm 4$ & $24 \pm 14$ & $28 \pm
6$ & $502 \pm 33$ & $13 \pm 21$ & \\
\multirow{2}{*}{\threel}
& Pre-fit & $126 \pm 31$ & $84 \pm 8$ & $23 \pm 3$ & $20 \pm 11$ & $24 \pm
5$ & $276 \pm 33$ & $32 \pm 6\phantom{0}$
& \multirow{2}{*}{258}\\
& Post-fit & $104 \pm 20$ & $84 \pm 8$ & $23 \pm 3$ & $19 \pm 10$ & $24
\pm 5$ & $254 \pm 18$ & $\phantom{0}7 \pm 11$ & \\
\hline
\multicolumn{10}{c}{\tHc}\\
\hline
\multirow{2}{*}{\twol}
& Pre-fit & $266 \pm 40$ & $165 \pm 19$ & $43 \pm 4$ & $25 \pm 15$ & $28 \pm
6$ & $526 \pm 39$ & $62 \pm 13$
& \multirow{2}{*}{514}\\
& Post-fit & $264 \pm 41$ & $165 \pm 18$ & $42 \pm 4$ & $20 \pm 11$ & $28 \pm 6$ & $520 \pm 36$ & $-3 \pm 25$ & \\
\multirow{2}{*}{\threel}
& Pre-fit & $126 \pm 31$ & $84 \pm 8$ & $23 \pm 3$ & $20 \pm 11$ & $24 \pm
5$ & $276 \pm 33$ & $30 \pm 6\phantom{0}$
& \multirow{2}{*}{258}\\
& Post-fit & $116 \pm 21$ & $84 \pm 8$ & $23 \pm 3$ & $15 \pm 8\phantom{0}$ & $23 \pm
5$ & $262 \pm 19$ & $-1 \pm 12$ & \\
\hline\hline
\end{tabular}
}
\end{center}
\end{table}
 
Binned maximum-likelihood fits to the distributions of the \twol
and \threel FCNC discriminants are performed to extract the best-fit values of
the \tHq branching fractions.  The profile likelihood technique is used, in
which systematic uncertainties are modeled as nuisance parameters
$\vec{\theta}$ which are
allowed to vary in the fit, constrained by Gaussian or log-normal probability
density penalty functions multiplying the likelihood function $\mathcal{L}$.  The test
statistic \qB is obtained from the profile log-likelihood ratio as $\qB \equiv -2 \ln
\Lambda_\mathcal{B} = -2 \ln \left[\mathcal{L}(\mathcal{B},
\hat{\hat{\vec{\theta}}})/\mathcal{L}(\mathcal{\hat B},
\hat{\vec{\theta}})\right]$, where $\mathcal{\hat B}$ and
$\hat{\vec{\theta}}$ are the \tHq branching fraction and nuisance parameter values
that give the global maximum likelihood and
$\hat{\hat{\vec{\theta}}}$ are the
nuisance parameter values which maximize $\mathcal{L}$ for a given branching
fraction $\mathcal{B}$.  The uncertainties in the best-fit branching fraction
value $\mathcal{\hat B}$ are determined by the variation of \qB by one unit
from its minimum, and the distribution of \qB is used to set 95\% CL upper
limits on the branching fractions \Br{\tHq} using the CL$_\mathrm{s}$ method \cite{cls}.  Due to
the near-degeneracy of the BDT response to \tHu and \tHc signals, during the fits,
one of these branching fractions is set to zero while the other is permitted
to float.  Expected and observed yields of events in the signal categories, before and after the nuisance
parameters are adjusted in the fits, are shown in Table~\ref{tbl:yields}. The distributions of the FCNC discriminant for the data and the best-fit signal-plus-background
models are shown in Figure~\ref{fig:fit}.
 
\begin{figure}
\begin{center}
\includegraphics[width=0.48\textwidth]{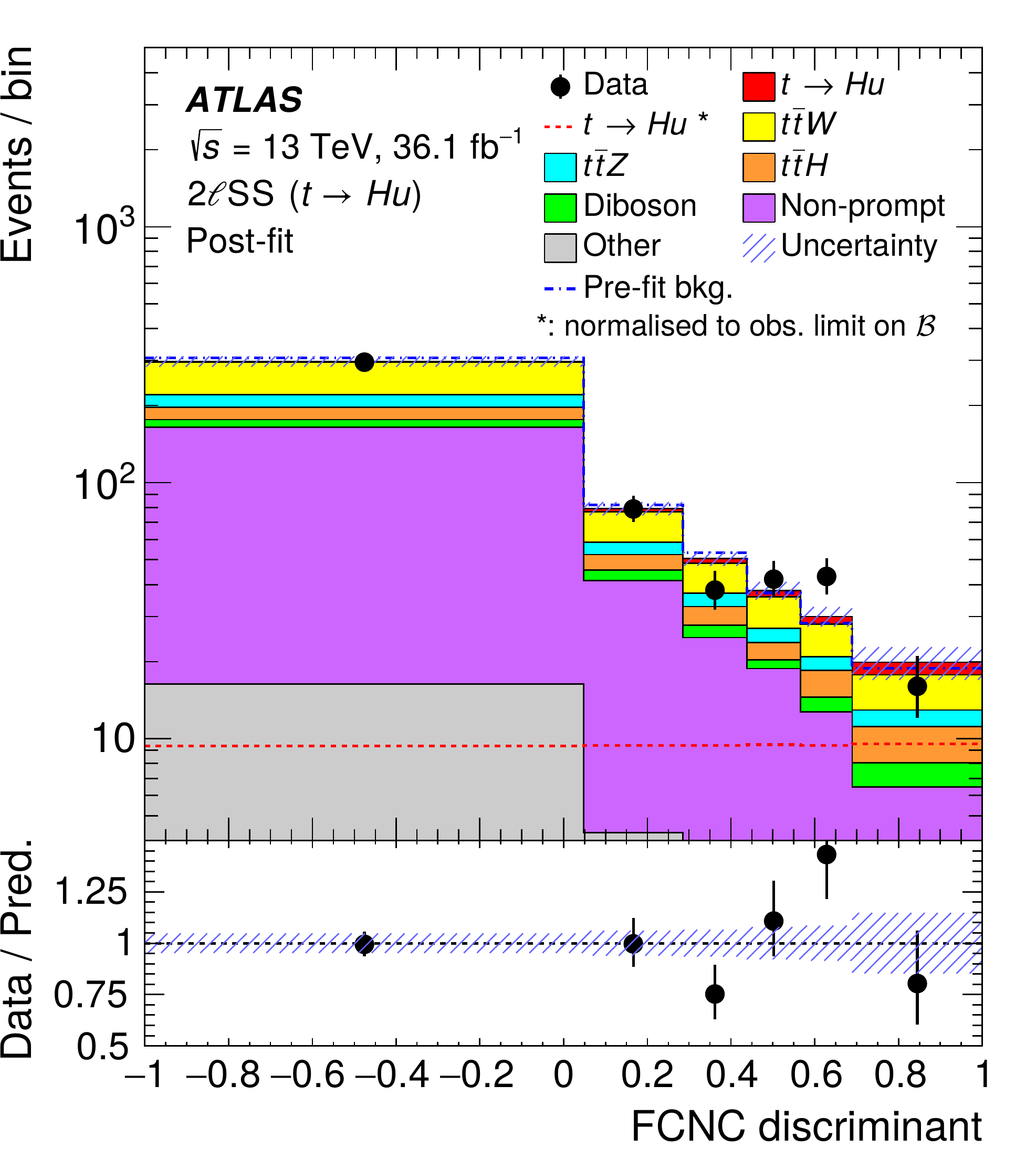}
\hfill\includegraphics[width=0.48\textwidth]{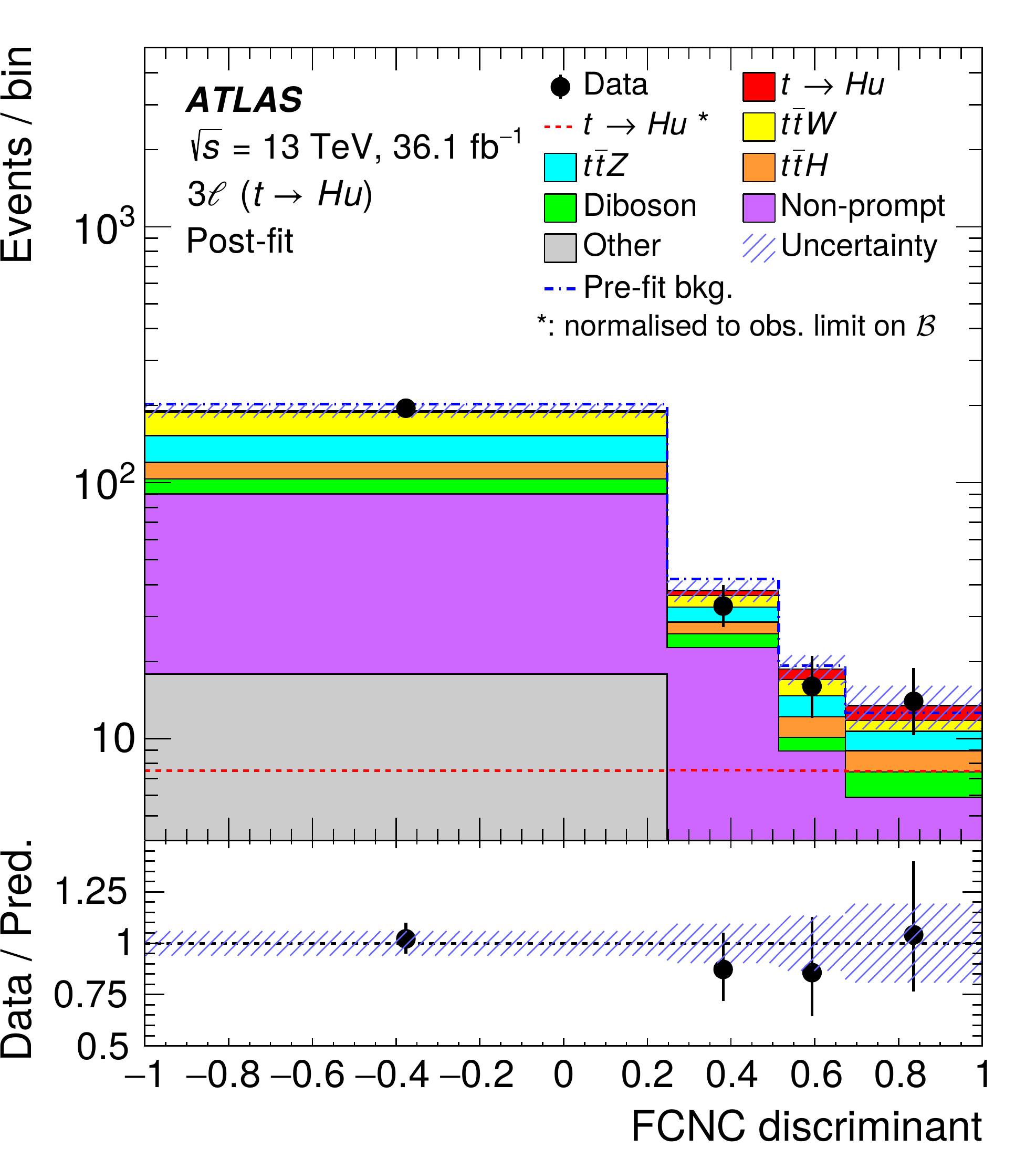} \\
\includegraphics[width=0.48\textwidth]{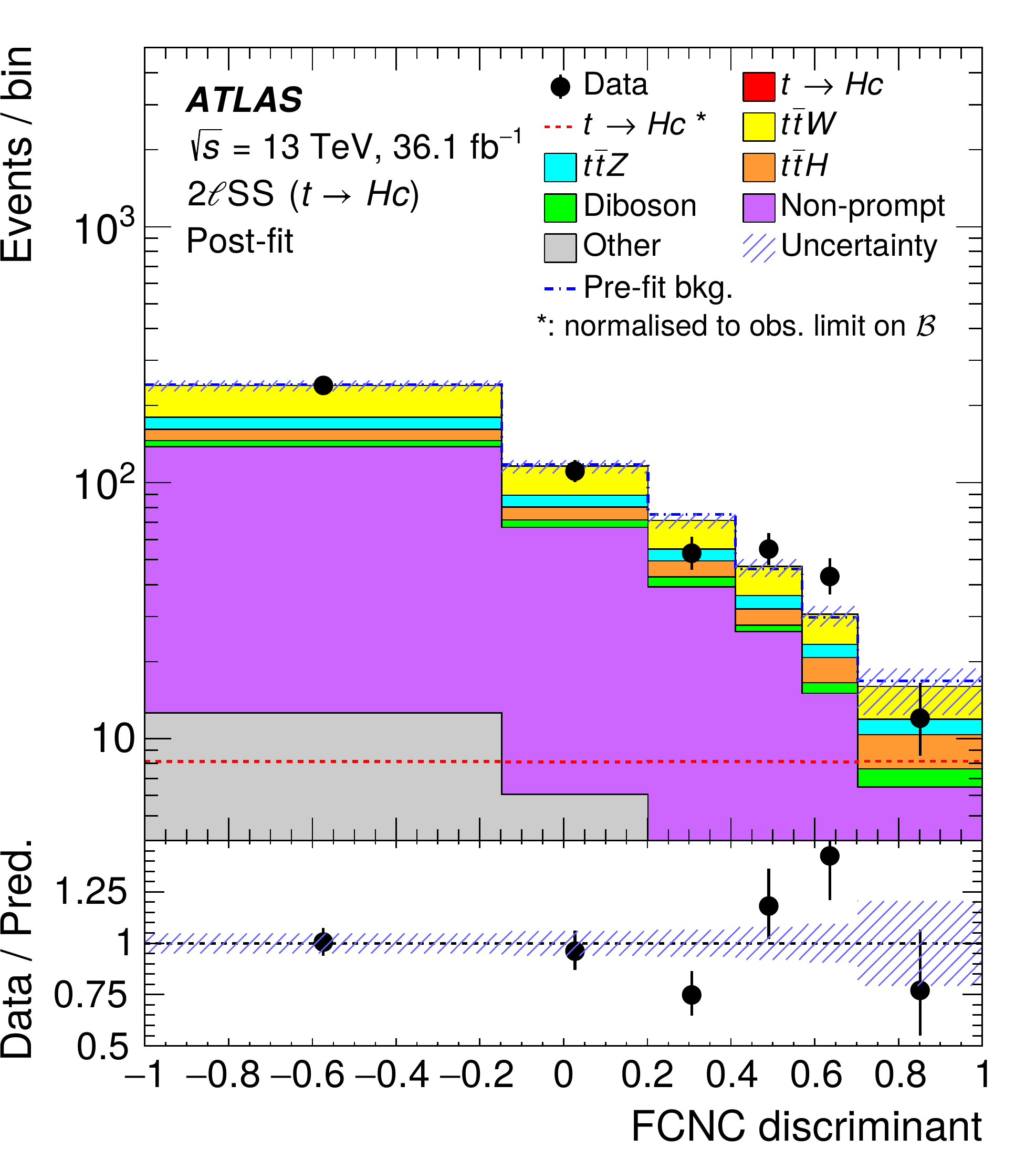}
\hfill\includegraphics[width=0.48\textwidth]{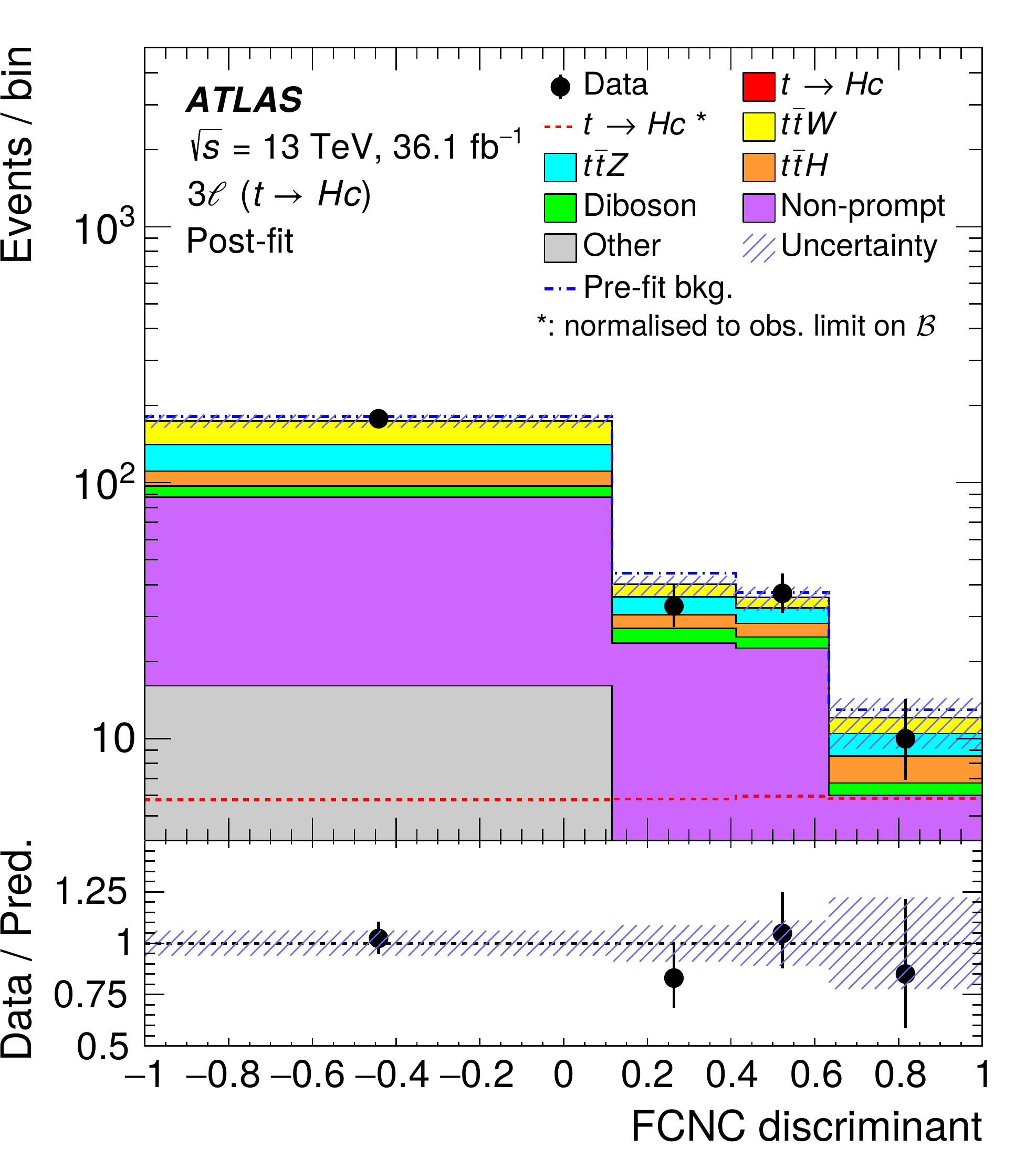}
\caption{\label{fig:fit}Distributions of the FCNC signal discriminants for
(top) \tHu, (bottom) \tHc signals in (left) \twol, (right) \threel
categories. In the \tHu fit, \Br{\tHc} is set to zero, and vice versa. The
binning, which is that used in the fit, is chosen so as to have roughly equal signal yields in each bin.  The
FCNC signals, normalized to their best-fit branching fractions from the
combined fit to the \twol and \threel categories, are shown as filled red histograms stacked above the
background components; only \tHu is large enough to be visible.  The hashed band indicates the total uncertainty in
the signal-plus-background prediction, including the statistical uncertainty
in the best-fit FCNC signal. The dashed red lines show the expected
contribution of the respective FCNC decay with the 95\% CL upper limit
branching fraction (0.19\% for \tHu, 0.16\% for \tHc). }
\end{center}
\end{figure}

\begin{table}
\renewcommand{\arraystretch}{1.4}
\begin{center}
\caption{\label{tbl:fitthu} Best-fit values and 95\% CL upper limits for
\Br{\tHu}, assuming $\Br{\tHc} = 0$. The ``stat.\ + syst.'' columns show
the full result allowing all systematic uncertainty nuisance parameters to float in the fit, while the
``stat.'' columns show the result with systematic uncertainty nuisance
parameters fixed to their values at the global best-fit point. }
\begin{tabular}{lcccc}
\hline\hline
& \multicolumn{2}{c}{Best-fit} & \multicolumn{2}{c}{Observed
(Expected)}\\
& \multicolumn{2}{c}{\Br{\tHu} [\%]} & \multicolumn{2}{c}{Upper Limit on
\Br{\tHu} [\%]}\\
\hline
& stat. & stat.\ + syst. & stat. & stat.\ + syst.\\
\hline
\twol & $\phantom{+}0.08\ ^{+0.08}_{-0.08}$ & $\phantom{+}0.08\ ^{+0.11}_{-0.10}$ & 0.23 (0.15) & 0.28
(0.21)\\
\threel & $\phantom{+}0.01\ ^{+0.09}_{-0.08}$ & $\phantom{+}0.01\ ^{+0.10}_{-0.09}$ & 0.20
(0.18) & 0.22 (0.21)\\
\hline
Combined & $\phantom{+}0.04\ ^{+0.06}_{-0.06}$ & $\phantom{+}0.04\ ^{+0.08}_{-0.07}$ & 0.17
(0.12) & 0.19 (0.15) \\
\hline\hline
\end{tabular}
\end{center}
\renewcommand{\arraystretch}{1.0}
\end{table}
 
\begin{table}
\renewcommand{\arraystretch}{1.4}
\begin{center}
\caption{\label{tbl:fitthc} Best-fit values and 95\% CL upper limits for
\Br{\tHc}, assuming $\Br{\tHu} = 0$. The ``stat.\ + syst.'' columns show
the full result allowing all systematic uncertainty nuisance parameters to float in the fit, while the
``stat.'' columns show the result with systematic uncertainty nuisance
parameters fixed to their values at the global best-fit point. }
\begin{tabular}{lcccc}
\hline\hline
& \multicolumn{2}{c}{Best-fit} & \multicolumn{2}{c}{Observed
(Expected)}\\
& \multicolumn{2}{c}{\Br{\tHc} [\%]} & \multicolumn{2}{c}{Upper Limit on
\Br{\tHc} [\%]}\\
\hline
& stat. & stat.\ + syst. & stat. & stat.\ + syst.\\
\hline
\twol & $\phantom{+}0.05\ ^{+0.08}_{-0.08}$ & $\phantom{+}0.05\ ^{+0.11}_{-0.10}$ & 0.22 (0.15) & 0.25
(0.20)\\
\threel & $-0.09\ ^{+0.10}_{-0.09}$ & $-0.09\ ^{+0.11}_{-0.11}$ & 0.19
(0.23) & 0.20 (0.25)\\
\hline
Combined & $-0.01\ ^{+0.06}_{-0.06}$ & $-0.01\ ^{+0.08}_{-0.08}$ & 0.15
(0.13) & 0.16 (0.15) \\
\hline\hline
\end{tabular}
\end{center}
\renewcommand{\arraystretch}{1.0}
\end{table}
 
\begin{figure}
\centering
\includegraphics[width=0.8\linewidth]{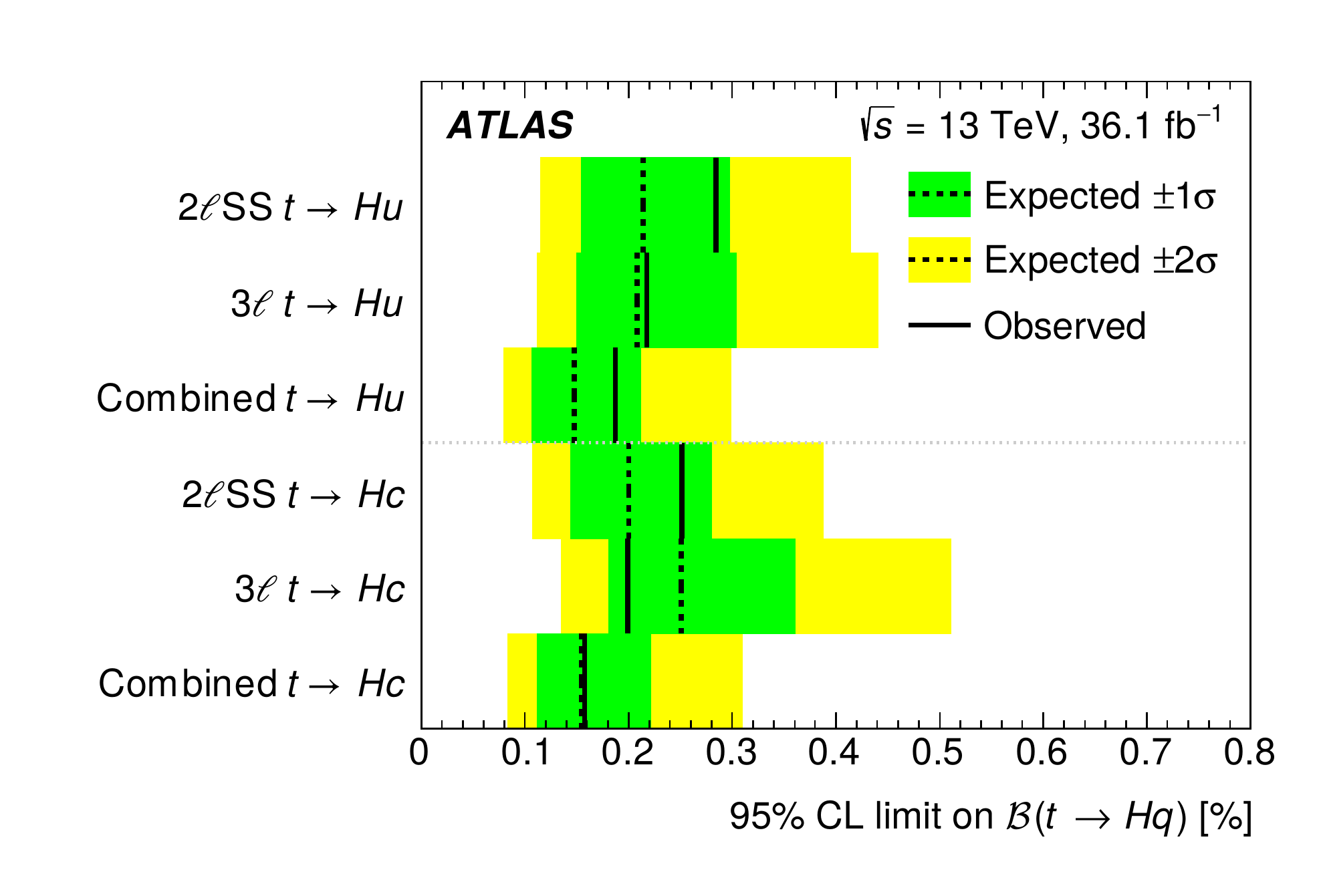}
\caption{\label{fig:limits}Observed and expected 95\% CL upper limits on \Br{\tHu}
and \Br{\tHc}. In each case, the other FCNC decay is assumed to have zero
branching fraction. The individual results from each signal category are shown as
well as the combination.}
\end{figure}
 
The results of the fits are shown in Tables~\ref{tbl:fitthu} and
\ref{tbl:fitthc}.  The best-fit branching fractions are compatible with zero, and 95\% CL
upper limits are set, as shown in Figure~\ref{fig:limits}.  Statistical uncertainties are dominant in the result.
No variations of the nuisance parameters by more than $1\sigma$ of the prior
systematic uncertainty are observed; the largest variations are observed in the
nuisance parameters associated with statistical uncertainties in the
non-prompt lepton background estimate and in one of the normalization
systematic uncertainties in
the \threel non-prompt lepton background.
 
To confirm the self-consistency of the non-prompt lepton background estimate,
a number of checks were performed. There is no evidence of a BDT response shape distortion
in the non-prompt lepton background estimate during the fit which would be
consistent with a FCNC signal.  In addition, a second fit that decorrelates
FCNC contributions in the signal categories from their impact on the non-prompt
lepton efficiency estimates finds best-fit branching fractions consistent with
those of the
nominal fit.


\section{Conclusion}
 
Any observable branching fraction for the FCNC decays \tHq would indicate new
physics beyond the Standard Model.  A search for \ttbar production events in which one top quark or
antiquark undergoes a \tHq decay was carried out with an integrated luminosity
of 36.1
\ifb\ of $pp$ collision data with $\sqrt{s} = 13$~\TeV\ collected in 2015 and
2016 using the ATLAS detector at the LHC.  Two final states are targeted: two same-charge light leptons with four or more jets, and three light leptons with two
or more jets. These are sensitive primarily to $H \to WW^*$ decays, with subleading
contributions from $H \to \tau\tau$ and $H \to ZZ^*$.  Specialized boosted decision trees using the kinematic properties of the final-state particles are used to distinguish FCNC signals from
non-prompt lepton backgrounds and from \ttW and \ttZ production.  Potential
contamination from FCNC signal in the non-prompt lepton background control
regions is treated in a self-consistent manner. No evidence of FCNC decays is
found and the upper limits set on the branching fractions are $\Br{\tHc} <$
0.16\% and $\Br{\tHu} <$ 0.19\% at 95\% CL.

\section*{Acknowledgments}
 
 
We thank CERN for the very successful operation of the LHC, as well as the
support staff from our institutions without whom ATLAS could not be
operated efficiently.
 
We acknowledge the support of ANPCyT, Argentina; YerPhI, Armenia; ARC, Australia; BMWFW and FWF, Austria; ANAS, Azerbaijan; SSTC, Belarus; CNPq and FAPESP, Brazil; NSERC, NRC and CFI, Canada; CERN; CONICYT, Chile; CAS, MOST and NSFC, China; COLCIENCIAS, Colombia; MSMT CR, MPO CR and VSC CR, Czech Republic; DNRF and DNSRC, Denmark; IN2P3-CNRS, CEA-DRF/IRFU, France; SRNSFG, Georgia; BMBF, HGF, and MPG, Germany; GSRT, Greece; RGC, Hong Kong SAR, China; ISF, I-CORE and Benoziyo Center, Israel; INFN, Italy; MEXT and JSPS, Japan; CNRST, Morocco; NWO, Netherlands; RCN, Norway; MNiSW and NCN, Poland; FCT, Portugal; MNE/IFA, Romania; MES of Russia and NRC KI, Russian Federation; JINR; MESTD, Serbia; MSSR, Slovakia; ARRS and MIZ\v{S}, Slovenia; DST/NRF, South Africa; MINECO, Spain; SRC and Wallenberg Foundation, Sweden; SERI, SNSF and Cantons of Bern and Geneva, Switzerland; MOST, Taiwan; TAEK, Turkey; STFC, United Kingdom; DOE and NSF, United States of America. In addition, individual groups and members have received support from BCKDF, the Canada Council, CANARIE, CRC, Compute Canada, FQRNT, and the Ontario Innovation Trust, Canada; EPLANET, ERC, ERDF, FP7, Horizon 2020 and Marie Sk{\l}odowska-Curie Actions, European Union; Investissements d'Avenir Labex and Idex, ANR, R{\'e}gion Auvergne and Fondation Partager le Savoir, France; DFG and AvH Foundation, Germany; Herakleitos, Thales and Aristeia programmes co-financed by EU-ESF and the Greek NSRF; BSF, GIF and Minerva, Israel; BRF, Norway; CERCA Programme Generalitat de Catalunya, Generalitat Valenciana, Spain; the Royal Society and Leverhulme Trust, United Kingdom.
 
The crucial computing support from all WLCG partners is acknowledged gratefully, in particular from CERN, the ATLAS Tier-1 facilities at TRIUMF (Canada), NDGF (Denmark, Norway, Sweden), CC-IN2P3 (France), KIT/GridKA (Germany), INFN-CNAF (Italy), NL-T1 (Netherlands), PIC (Spain), ASGC (Taiwan), RAL (UK) and BNL (USA), the Tier-2 facilities worldwide and large non-WLCG resource providers. Major contributors of computing resources are listed in Ref.~\cite{ATL-GEN-PUB-2016-002}.
 

\clearpage

\printbibliography
 
\clearpage
\input{atlas_authlist}

\end{document}

%% file: atlas_authlist.tex
 
\begin{flushleft}
{\Large The ATLAS Collaboration}

\bigskip

M.~Aaboud$^\textrm{\scriptsize 34d}$,    
G.~Aad$^\textrm{\scriptsize 99}$,    
B.~Abbott$^\textrm{\scriptsize 124}$,    
O.~Abdinov$^\textrm{\scriptsize 13,*}$,    
B.~Abeloos$^\textrm{\scriptsize 128}$,    
D.K.~Abhayasinghe$^\textrm{\scriptsize 91}$,    
S.H.~Abidi$^\textrm{\scriptsize 164}$,    
O.S.~AbouZeid$^\textrm{\scriptsize 143}$,    
N.L.~Abraham$^\textrm{\scriptsize 153}$,    
H.~Abramowicz$^\textrm{\scriptsize 158}$
% ...
\bigskip
\\

% ...
$^{13}$Institute of Physics, Azerbaijan Academy of Sciences, Baku; Azerbaijan.\\
% ...
$^{34}$$^{(a)}$Facult\'e des Sciences Ain Chock, R\'eseau Universitaire de Physique des Hautes Energies - Universit\'e Hassan II, Casablanca;$^{(b)}$Centre National de l'Energie des Sciences Techniques Nucleaires (CNESTEN), Rabat;$^{(c)}$Facult\'e des Sciences Semlalia, Universit\'e Cadi Ayyad, LPHEA-Marrakech;$^{(d)}$Facult\'e des Sciences, Universit\'e Mohamed Premier and LPTPM, Oujda;$^{(e)}$Facult\'e des sciences, Universit\'e Mohammed V, Rabat; Morocco.\\
% ...
$^{91}$Department of Physics, Royal Holloway University of London, Egham; United Kingdom.\\
% ...
$^{99}$CPPM, Aix-Marseille Universit\'e, CNRS/IN2P3, Marseille; France.\\
% ...
$^{124}$Homer L. Dodge Department of Physics and Astronomy, University of Oklahoma, Norman OK; United States of America.\\
% ...
$^{128}$LAL, Universit\'e Paris-Sud, CNRS/IN2P3, Universit\'e Paris-Saclay, Orsay; France.\\
% ...
$^{143}$Santa Cruz Institute for Particle Physics, University of California Santa Cruz, Santa Cruz CA; United States of America.\\
% ...
$^{153}$Department of Physics and Astronomy, University of Sussex, Brighton; United Kingdom.\\
% ...
$^{158}$Raymond and Beverly Sackler School of Physics and Astronomy, Tel Aviv University, Tel Aviv; Israel.\\
% ...
$^{164}$Department of Physics, University of Toronto, Toronto ON; Canada.\\
% ...
$^{a}$ Also at  Department of Physics, University of Malaya, Kuala Lumpur; Malaysia.\\
$^{b}$ Also at Borough of Manhattan Community College, City University of New York, NY; United States of America.\\
$^{c}$ Also at California State University, East Bay; United States of America.\\
$^{d}$ Also at Centre for High Performance Computing, CSIR Campus, Rosebank, Cape Town; South Africa.\\
$^{e}$ Also at CERN, Geneva; Switzerland.\\
% ...
$^{*}$ Deceased

\end{flushleft}
